\documentclass[epj]{webofc}
\usepackage[varg]{txfonts}   
\def\be{\begin{equation}}
\def\ee{\end{equation}}
\woctitle{CNR15}
\begin{document}
\title{
A new approach to barrier-top fission dynamics
}

\author{G.F.~Bertsch\thanks{\email{bertsch@uw.edu}} and J.M. Mehlhaff 
}

\institute{
Department of Physics and Institute for Nuclear Theory,\\
University of
Washington, Seattle, WA USA
          }

\abstract{%
We proposed a calculational framework for describing induced fission
that avoids the Bohr-Wheeler assumption of well-defined fission 
channels.
The building blocks of our approach are configurations that form a
discrete, orthogonal basis and can be characterized by both energy
and shape.
The dynamics is to be
determined by interaction matrix elements between the states rather than by
a Hill-Wheeler construction of a collective coordinate. Within our approach,
several simple limits can be seen: diffusion; quantized conductance;
and ordinary decay through channels.  The specific proposal for the
discrete basis is to use the $K^\pi$ quantum numbers of the axially
symmetric Hartree-Fock approximation to generate the configurations.
Fission paths would be determined by hopping from configuration
to configuration via the residual interaction.
We show as an example the configurations needed to describe a
fictitious fission decay $^{32}{\rm S} \rightarrow ^{16}{\rm O}
+ ^{16}{\rm O}$.  We also examine the geometry of the
path for fission of $^{236}$U, measuring distances by the number
of jumps needed to go to a new $K^\pi$ partition.
}
\maketitle
\section{Introduction}
\label{intro}
\begin{figure}[tb]
\centering
\includegraphics[width=11cm]{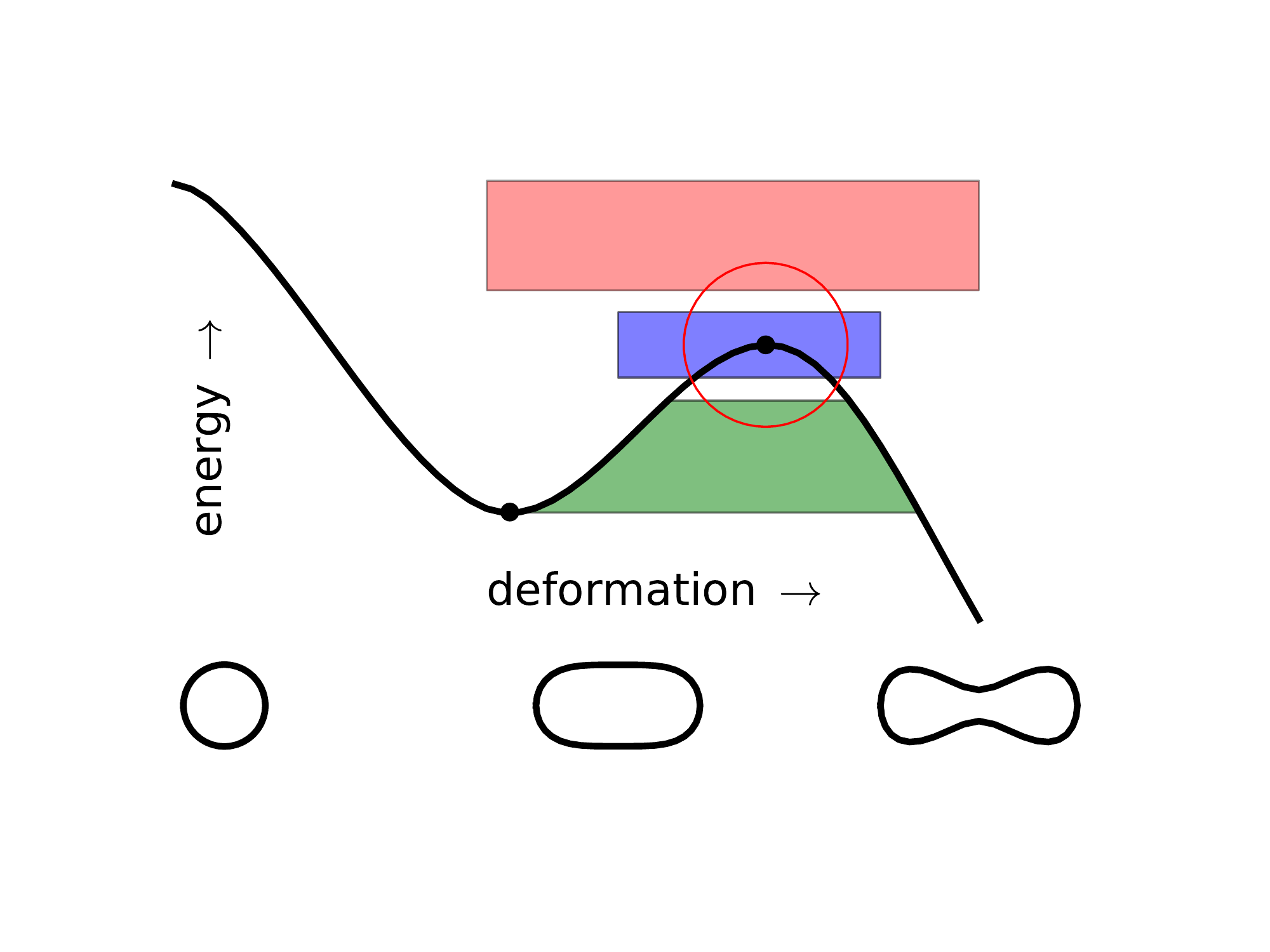}
\caption{\label{fig0} Schematic view of the fission landscape
with a simple barrier.  The colored regions show the areas
of different dynamics: diffusive in pink, tunneling in green,
and barrier-top in blue.  
}
\end{figure}

In this talk we will advocate a radically different approach to 
the theory of induced fission.  To put this into context, we
show in Fig. \ref{fig0} a schematic view of the fission landscape 
with the different energy regions indicated by color.
In the tunneling region, shown in green, the dynamics is driven
by the pairing interaction.  High above the barrier, shown in pink,
we expect the dynamics to be highly overdamped and amenable to
treatment by a diffusion equation.  In between is the barrier-top
region, shown in blue.  This  commonly described
by the Bohr-Wheeler theory and its generalization to multiple
barriers.  

We seek an alternate treatment that does not require
well-behaved channels to cross the barrier.
The concept of a channel demands that the wave function can
be written as a product of an internal part and a one-dimensional
function of some collective coordinates.  This is leads to the
Born-Oppenheimer approximation
in molecular dynamics, which is well justified by the large 
separation of electron and nucleus mass scales.  But this
not at all the case in nuclear physics, where both single-particle
and collective structures are on the same 1 MeV energy scale. 
The tools for microscopic
calculations in that framework are just not up to the task.  

\begin{figure}[tb]
\centering
\includegraphics[width=13cm]{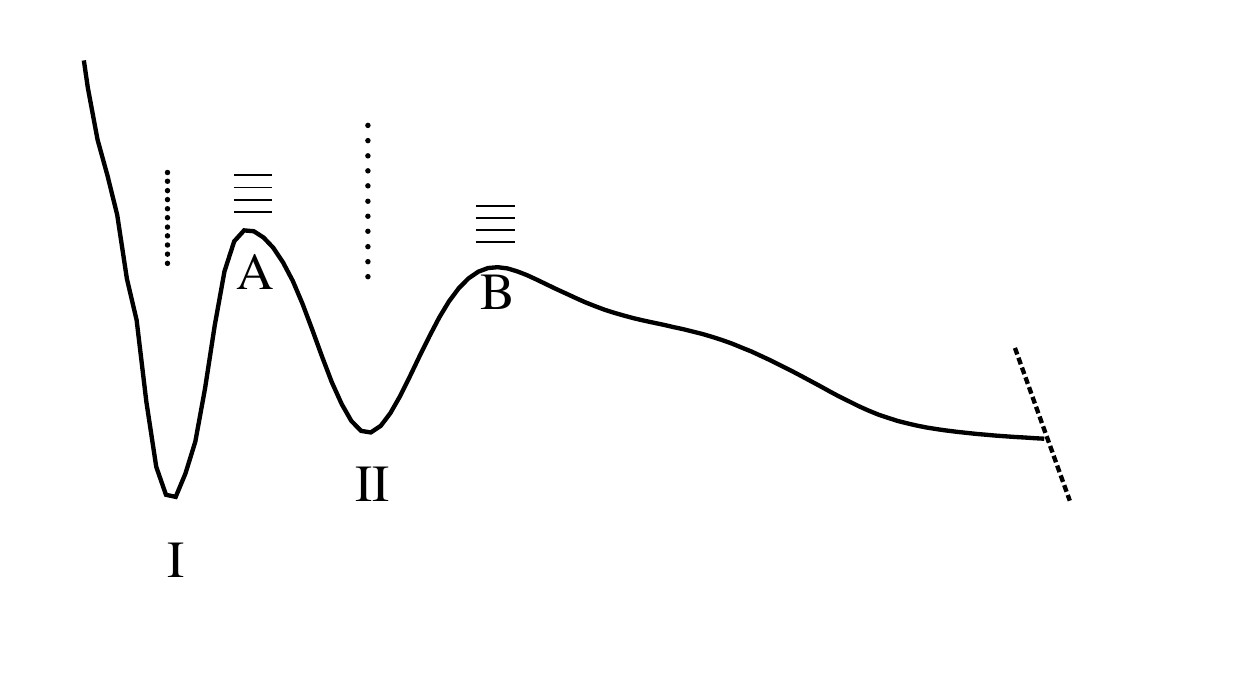}
\caption{\label{fig1a}
Fission potential energy surface (PES) for $^{236}$U, taken from Fig. 7 of Ref. \cite{ro14}.  Dashed
line shows energy of separated fission fragments at the same quadrupole
deformation coordinate.  Superimposed on the PES are the elements needed
for the generalized Bohr-Wheeler description of the dynamics. 
Type I and II state are indicated by dots;
channels bridging the barriers are indicated by horizontal lines.
In practice, the channels are only needed at the barrier tops.
}
\end{figure}

The present state of the art for modeling induced fission may be
seen in the recent calculations of the Los Alamos group \cite{bo13,bo14}.
The basic calculational framework was laid out by Bj\o rnholm
and Lynn \cite{bj80}, who generalized the Bohr-Wheeler
theory to deal with a fission landscape having two barriers.  The degrees of
freedom are depicted in Fig. \ref{fig1a}.  
They are the state densities 
in wells I and II, indicated by dots in the Figure, and the
channels at barriers A and B, indicated by the horizontal lines.
The final formula for the fission decay rate is a generalization
of the Bohr-Wheeler decay rate formula
\be
W = { 1\over 2 \pi \hbar \rho} \sum_c T_c
\ee 
which only has only one compound nucleus density $\rho$ and only 
transmission coefficients $T_c$ for a single barrier.

To connect this to a microscopic theory based on a nucleonic 
Hamiltonian, we need to know how to calculate the transmission coefficients
$T_c$ 
between states and channels,
and also   
how calculate mixing between channels.  The microscopic
framework used up to now is the Generator
Coordinate Method (GCM).  Unfortunately, it does not have a natural place
for the ordinary states and the needed connection to channels
seems difficult to incorporate.  For the mixing
of channels, there is a heroic
attempt by Gogny's group to set up a GCM framework for this
purpose \cite{be11},
but it appears to us to be very complicated to carry out in practice.

In our view, the problem is the channel picture itself.  Channels are
useful if there is clear separation between collective and intrinsic 
degrees of freedom.  That is O.K. for the Born-Oppenheimer 
framework for molecular physics, but the separation of collective and
intrinsic energy scales is simply not present in nuclear
physics.  Another problem is the non-orthogonality of the channel
wave functions in the GCM.
In particular, the over-completeness of the basis often
gives rise numerical stabilities that can only be suppressed by ad
hoc truncations.
 
From the phenomenological side, the dynamics of induced fission
may be much closer to a diffusive limit than an inertial 
limit.  One sees from many studies including one presented 
in this Workshop \cite{le15} and a recent one on mass distributions
\cite{ra11} that  statistical and diffusive models can
describe many features of the fission final state.  Furthermore,
we also heard  in the Workshop a report on a microscopic
dynamical model that produced a fission time so large that inertial
motion would be highly over-damped \cite{bu15}.

\section{Dynamics in a discrete basis}
\label{sec-1}

There is an alternative.  That is to construct a discrete basis for
the Hamiltonian, avoiding completely the introduction of 
continuous collective degrees of freedom.  In this section we
show how various limits of the dynamics can emerge, deferring to
the next section how we envisage constructing the basis.  The basis
will be composed of 
mean-field configurations, allowing one to calculate with
well-known methods the matrix elements of Hamiltonians of the
usual microscopic form,
\be
\label{H}
H = \sum \varepsilon_i a^\dagger_i a_i +\frac{1}{4}\sum
v_{ij,kl} a^\dagger_i a^\dagger_j a_l a_k.
\ee
Besides the close connection to
configuration-interaction (CI) computational methods that have been
so successful with the nuclear shell model, the discrete-basis
framework provides a conceptual bridge to quantum transport in condensed
matter physics.

The states need to be characterized by energy and some measure 
of the shape; these may be 
determined by the expectation values of the Hamiltonian and some
single-particle operator such as the quadrupole moment.  It is
then easy to set up conditions on Hamiltonian to realize different
possibilities for the dynamics.  

\begin{figure}[tb]
\includegraphics[width=7cm]{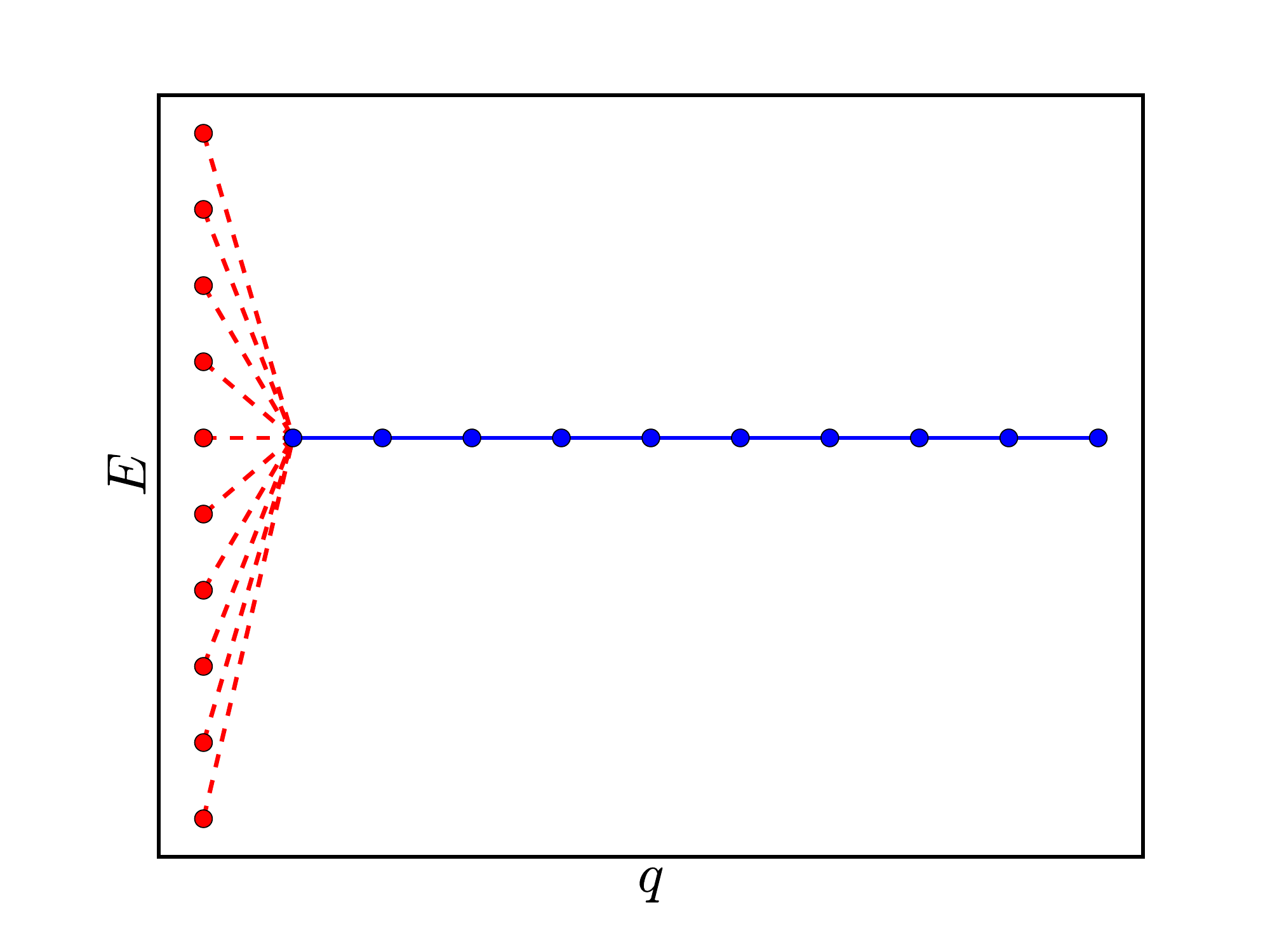}%
\includegraphics[width=6cm]{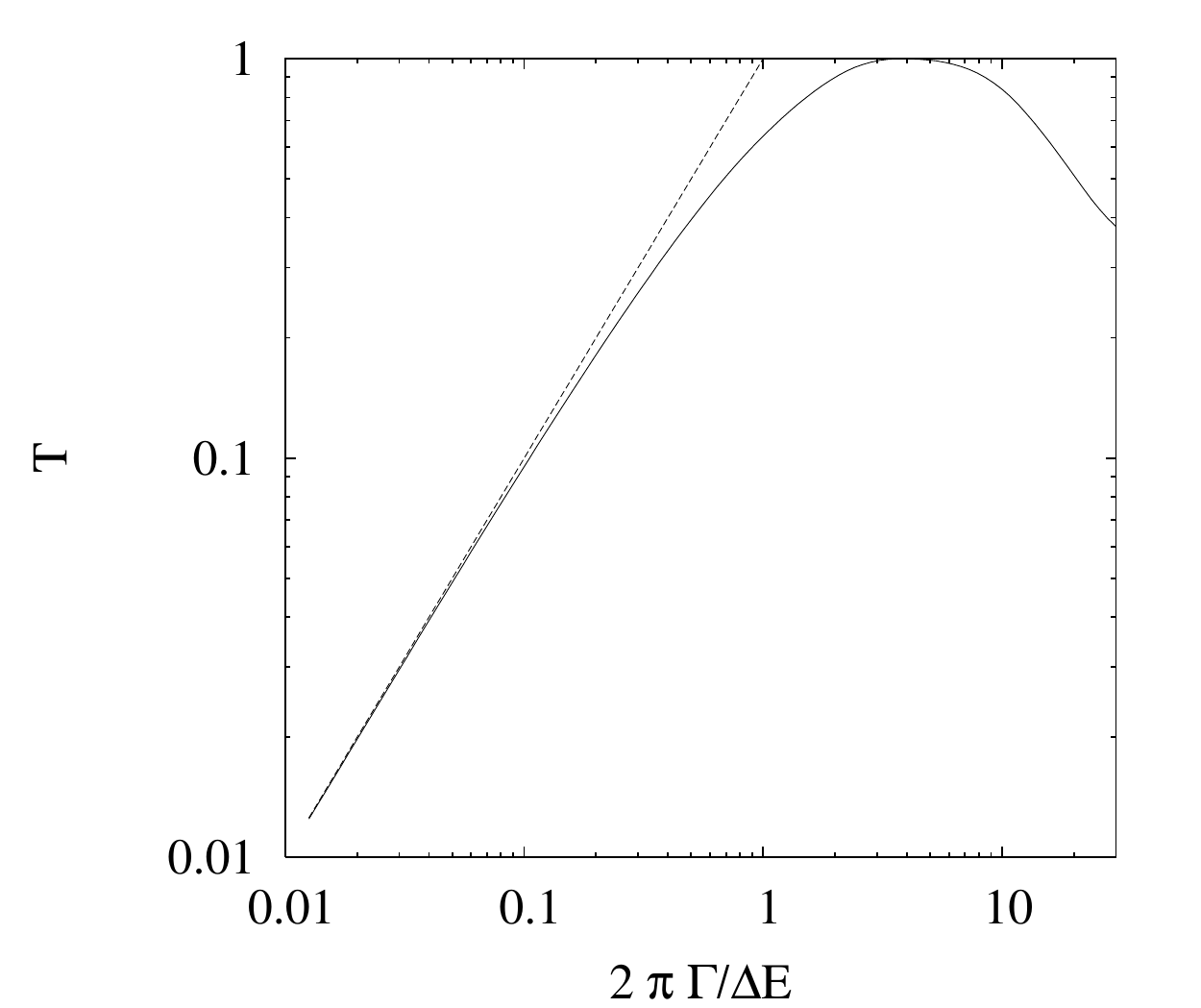}
\caption{\label{single-channel}  Discrete-state modeling of the compound nucleus
coupling to a neutron decay channel. On the left are shown the
states as circles and their couplings as lines.  On the right is
shown the calculated transmission coefficient $T$ compared to the
weak-absorption limit $T = 4 \pi^2 v_c^2 \rho^2$.  
}
\end{figure}

The first limit is that of a compound nucleus decaying into a 
single channel.  The basis states are set up as shown in 
Fig.~\ref{single-channel} on the left.
The compound nucleus states are shown as the tower of states
indicated as red points.  The channel states are the regularly
spaced blue point on the left.  The interactions between the
compound states and the first channel state are shown as the
dashed red lines.  They would be taken as Porter-Thomas 
distributed about some rms average value $\langle v^2_c \rangle$.
The interactions between channel states are
shown as the solid blue lines.   They would all be equal.  
Then the entire physics of scattering theory and compound
nucleus resonances can be displayed by using standard methods.
There has been recent interest in couplings to the continuum
that violate Porter-Thomas statistics \cite{mi15}; perhaps this model would
be useful to explore such possibilities.

\begin{figure}[tb]
\includegraphics[width=7cm]{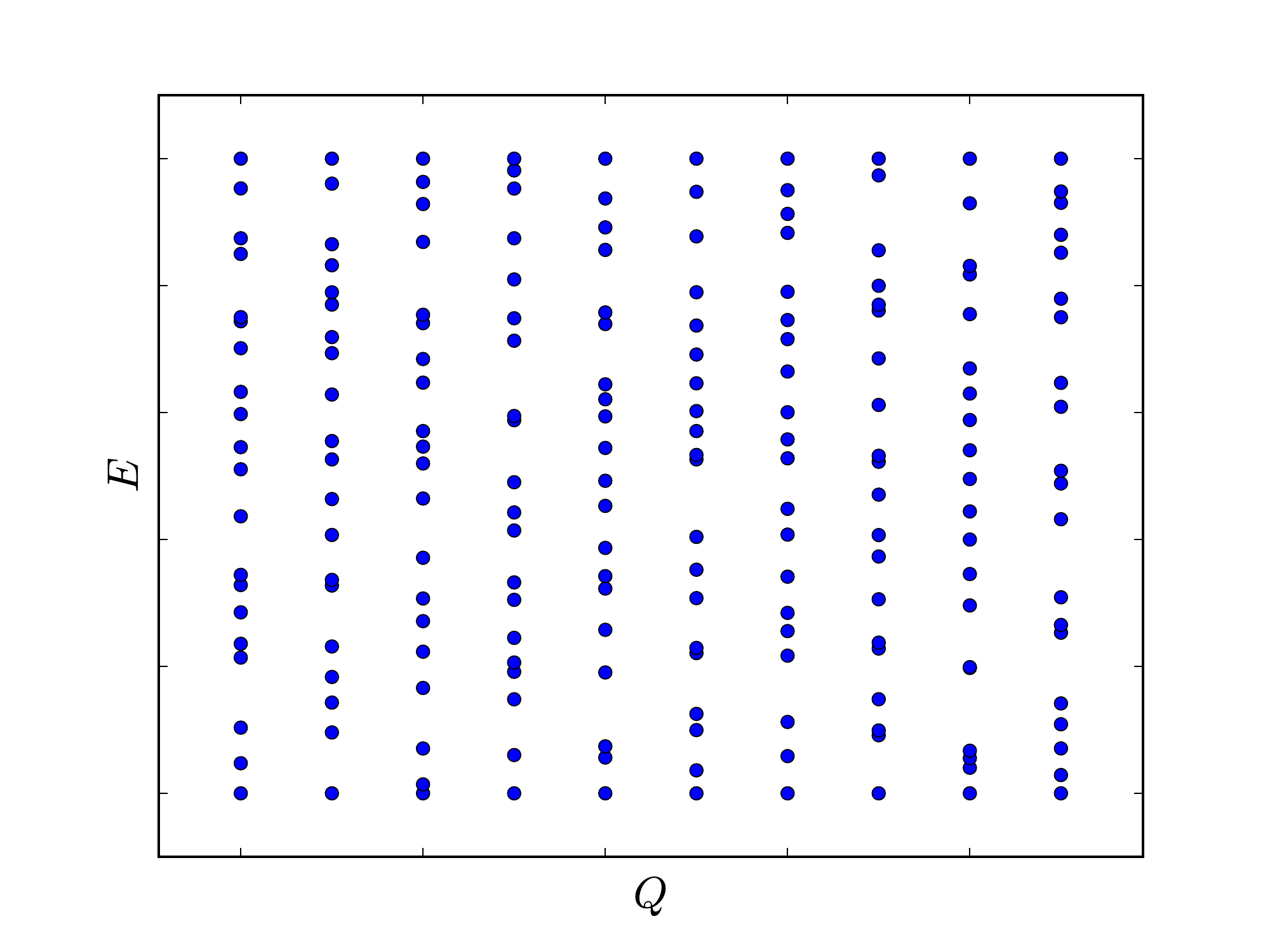}
\includegraphics[width=7cm]{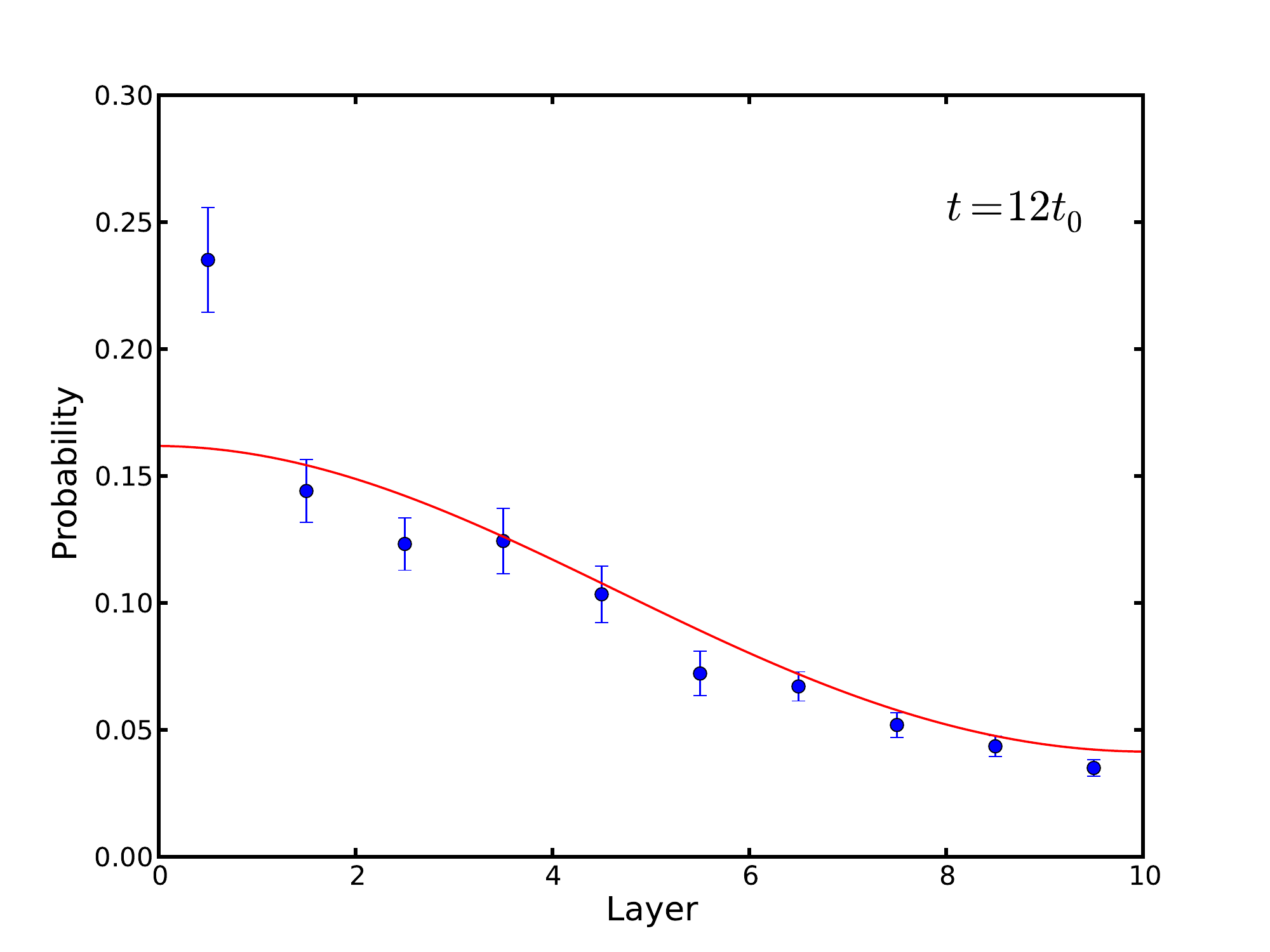}
\caption{\label{diffusion}  Discrete basis modeling of shape diffusion
in a region of high level density.  The states are grouped in vertical
layers by 
their deformation, and it is assumed that only nearby layers
are coupled by the interaction.  The right hand side shown 
the probability distribution after some time interval, starting
from an initial wave function localized on the first layer.
}
\end{figure}

Another limit is one that approximates diffusive dynamics.  By
this we mean that wave packets decay according to
the diffusion equation with some diffusion coefficient $D(q)$ according
to the equation
\be
{\partial P \over \partial t} = {\partial \over \partial q}D(q) 
{\partial P\over \partial q}
\ee
where $q$ is a shape index. To realize this limit in a discrete
basis, we distribute the states in layers according to the
shape parameter $q$ and in energy according to 
Gaussian random matrix ensemble.
The layout of states in the $(E,q)$ plane is shown in 
Fig. \ref{diffusion} on the left.
We also assume a Porter-Thomas distribution of interaction matrix
elements, limited to states on next-neighbor layers.
Evolving the dynamics by the time-dependent Schr\"odinger, we find
probability distributions on the different layers at later times.
The results are shown on the right-hand panel of the Figure at
some fixed time.  The points
are an average of different Hamiltonians in the ensemble, with the
error bars showing the r.m.s. fluctuations.
The connection to macroscopic theory comes if we can relate
the statistical properties of the Hamiltonian ensemble to 
a diffusion coefficent.
This is 
given by the formula \cite{bu92}
\be
\label{D}
D = 2 \pi\rho(E) \overline{ (q_\alpha - q_\beta)^2\langle \alpha | v|
\beta\rangle^2}
\ee
where $\rho$ is the level density in a single layer and the bar 
indicates an average over interactions between states in different
layers.
The red curve is the corresponding distribution predicted by the
diffusion equation, the $D$ from  Eq. (\ref{D}).
One sees that the classical physics is quite accurate except
for the some trapping in first layer.

\begin{figure}[tb]
\includegraphics[width=7cm]{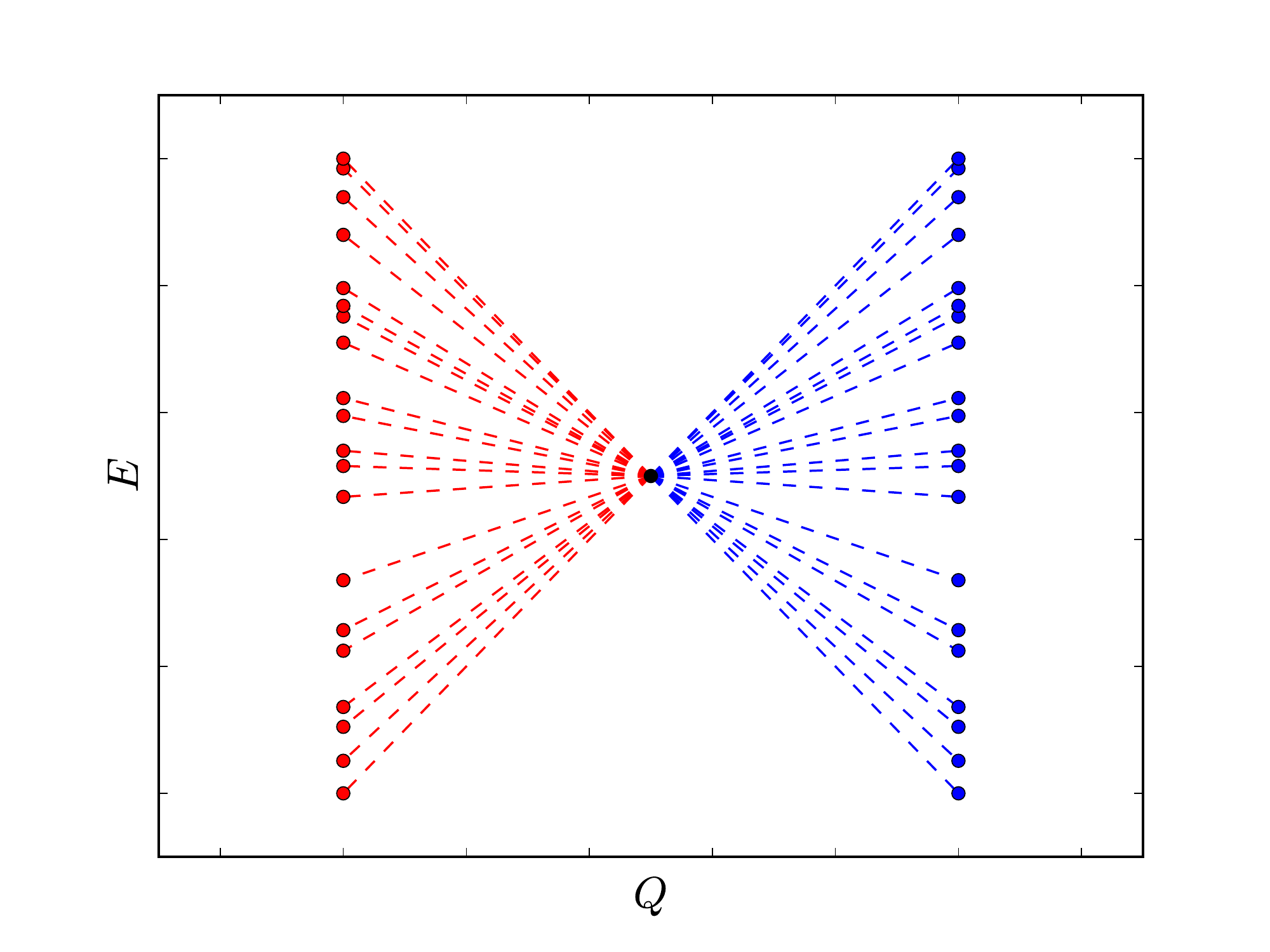}%
\includegraphics[width=8cm, trim =15 0 10 0, clip]{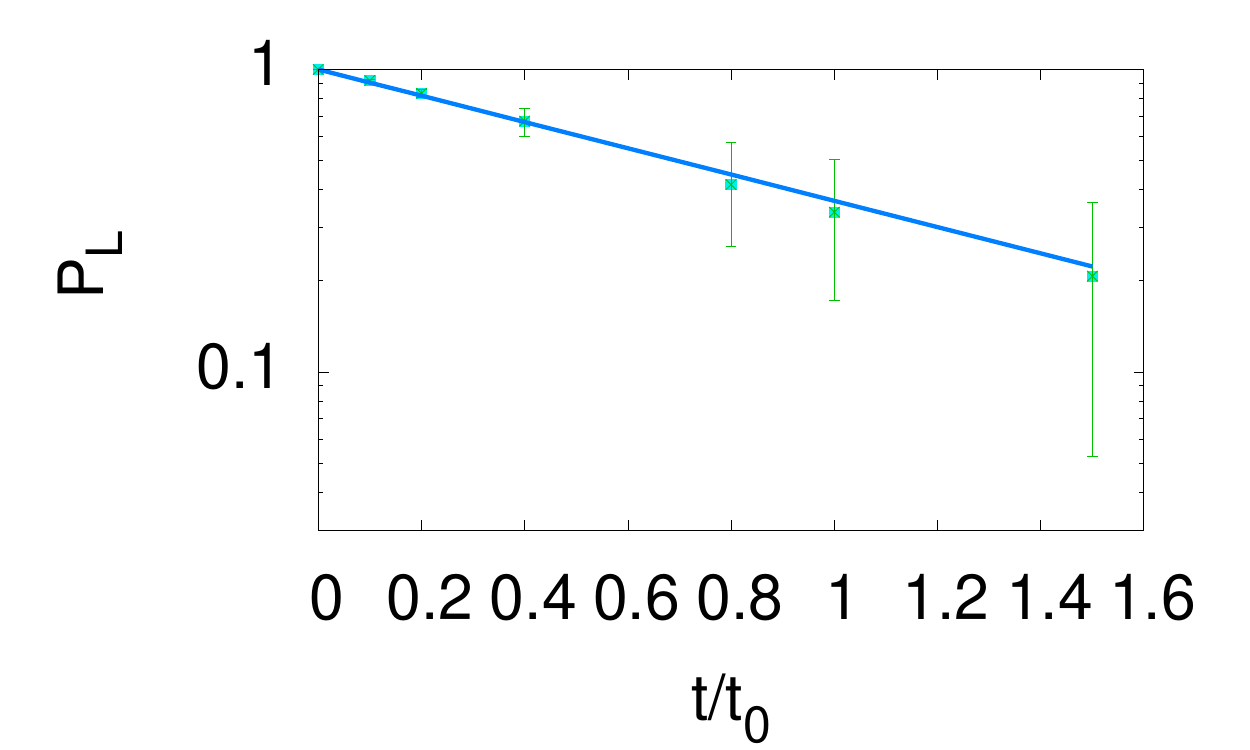}%
\caption{\label{quantum-dot}  Discrete state representation of 
resonance-mediated conductances.  The states of the compound nucleus
are on the left, and the post-barrier states are on the right.
}
\end{figure}

Another interesting limit is what may be called resonance-mediated
conductance.  It is inspired as a very simplified model for
fission dynamics at the barrier top as well as for the conduction
of electrons from one conductor to another through a quantum
dot.  The layout of states is shown in Fig. \ref{quantum-dot}.  
The transport
takes place through the single state in the middle.  There
is a lot of physics that can be explored with this model \cite{be14}.
For this talk, we just mention one limit.  With suitable restrictions
on level densities and interaction strengths, 
the rate at
which the system goes from one side to the other can be
calculated by Eq. (1) with the transmission coefficient given
by the formula \cite{al00}
\be
\label{Tr}
T_r  =  {\Gamma_{R} \Gamma_{L} \over E_b^2 + (\Gamma_{R}+\Gamma_{L})^2/4}.
\ee
Here the decay widths of the resonance to the left and to the right
($\Gamma_L$ and $\Gamma_R$) are calculated by Fermi's
Golden Rule.  Again, we examine an ensemble of Hamiltonians using
random matrix energy spacings, but now with fixed couplings to the
resonance to satisfy Eq. (\ref{Tr}) with $T_r=1$.  The resulting time-dependent probability for staying
on the left is shown in the panel on the right, with error bars
showing the fluctuations in the ensemble.  The solid line is 
an exponential decay with the decay rate determined by Eq. (1) with
the calculated transmission factor.  We see that the average decay
rate is well reproduced, but there is considerable fluctuation due
to the level-density fluctuations in the random-matrix ensemble.

\section{An implementation: the axial basis}
\label{sec-2}
A discrete basis should be composed of orthogonal states, and it
should be extendible in principle to a complete basis.
A good candidate to satisfy these requirements is what we shall call the 
axial basis.  As in traditional theory, the underlying framework is 
self-consistent mean-field theory.  But
instead of adding a generator-coordinate field to distinguish states,
we use the partition of the particle numbers to orbitals of different
K quantum numbers to make a first landscape of the PES.  To make this
clear, let us consider a very simple example, the shell-model ground
state of $^{16}$O.  The filled shells are $s_{1/2},p_{3/2},p_{1/2}$.
The $K$ quantum number can be taken as the azimuthal angular momenta
$j_z$ of the shells.  This produce the fillings shown in Table I,
amounting to a partition of the 16 nucleus into 6 $K^{\pi}$ sets.
\begin{table}
\centering
\caption{Orbital filling for $^{16}$O in the spherical shell model.
The columns separate the $K$ quantum number, and the rows separate
nucleon type and parity.
}
\label{tab-1}       
\begin{tabular}{|c|ccc|}
\hline
$K$  & $1/2$& $3/2$& $5/2$\\
\hline
p$^+$& 2 &0&0\\
p$^-$& 4 & 2 &0\\
n$^+$& 2 &0&0\\
n$^-$& 4 &2 &0\\
\hline
\end{tabular}
\end{table}

\begin{figure}[tb]
\centering
\includegraphics[width=12cm]{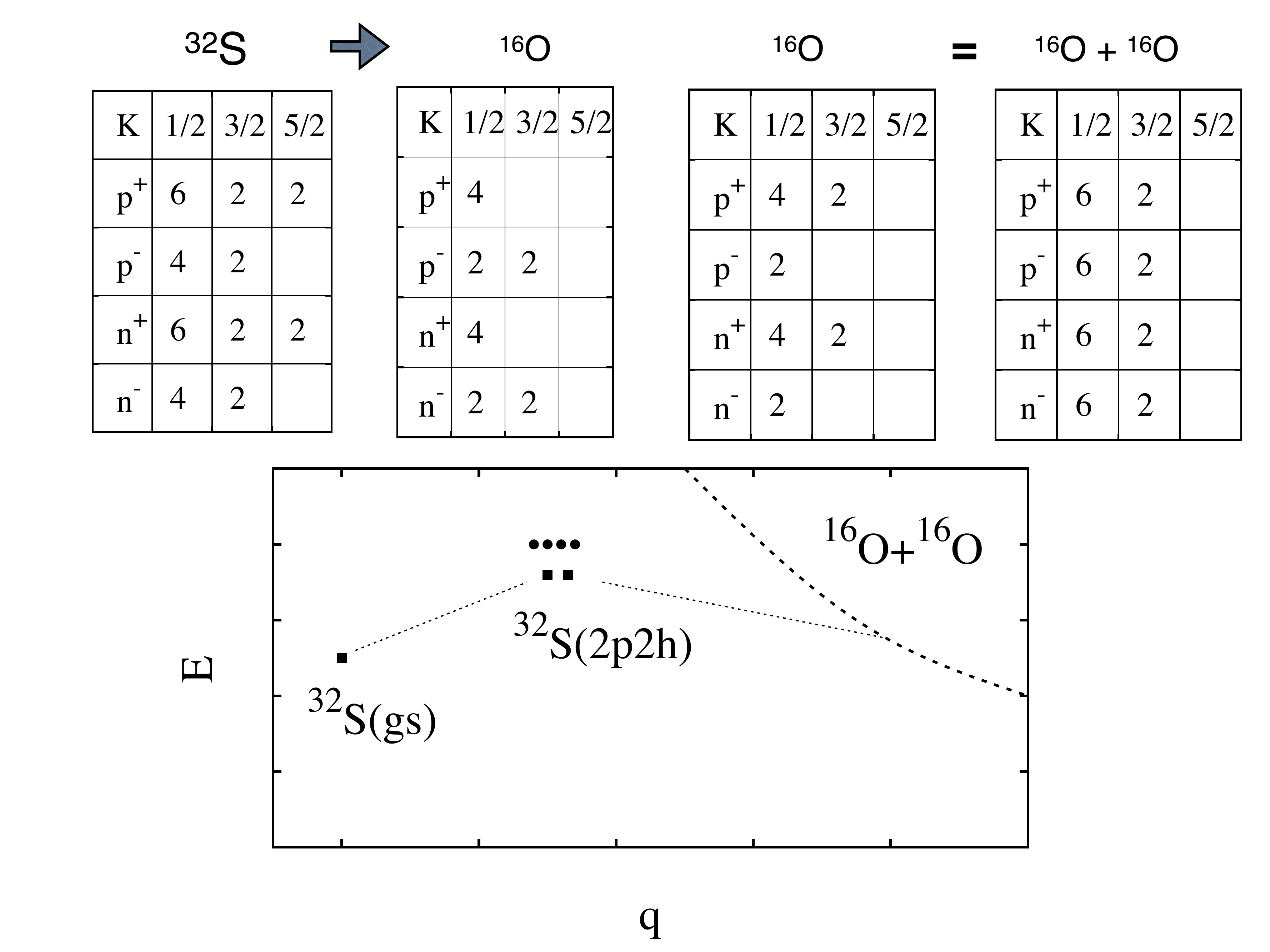}
\caption{\label{s32}  Fission of $^{32}S$ in a toy model.  
See text for explanation.
}
\end{figure}

To see how the scheme might work, we examine the partitions for a toy model,
the fission of $^{32}$S into two $^{16}$O nuclei with a Hamiltonian tuned to
allow the decay.  This model was proposed
in Ref.~\cite{negele}.  Tables of 
the partitions are shown in Fig. \ref{s32}.  
The leftmost table gives the $K^\pi$ partition
for $^{32}$S as a spherical shell-model configuration.  The filled
shells are $0s_{1/2},p_{3/2},p_{1/2},d_{5/2},1s_{1/2}$.  Notice that
the nonzero $K^\pi$ orbitals go up to the $5/2^+$ associated with
the $d_{5/2}$ shell.  The partition table for $^{16}$O+$^{16}$O can be
constructed by a very simple argument.  The orbitals of the combined
system have the same $K$ quantum numbers as the individual nuclei if
the fission is along the $z$-axis.  Each orbital has an extra two-fold
degeneracy because of its presence in both O nuclei. The plus and
minus linear combinations will have good parity.  Thus, each orbital
in the individual O gives rise to two orbitals with the same $K$ 
but opposite parities in the combined system.  The resulting partitions
are shown on the right hand side of the Figure.  Comparing the
initial and final partitions, one sees that the decay requires
four particles to be moved from $K^\pi = 5/2^+$ orbitals to 
$K^\pi= 1/2^-$ orbitals.  The various configurations involved
in the decay are depicted in the energy-versus-shape graph in the
bottom panel of the figure.  On the left is the $^{32}$S ground state.
The residual interaction connects it (dotted line) with several 2-particle
2-hole states as depicted in the middle.  Finally, a second application
of the residual interaction connects the intermediate configurations
to the final partition.  We have indicated the last configuration with a dashed 
line because it is not clear whether it would have a stable minimum
in the Hartree-Fock minimization.  It might be that the energy decreases
continuously as the distance between centers increases.  At that
point we cannot avoid dealing with the problem of coupling to the
continuum.  

\begin{figure}[tb]
\centering
\includegraphics[width=11cm]{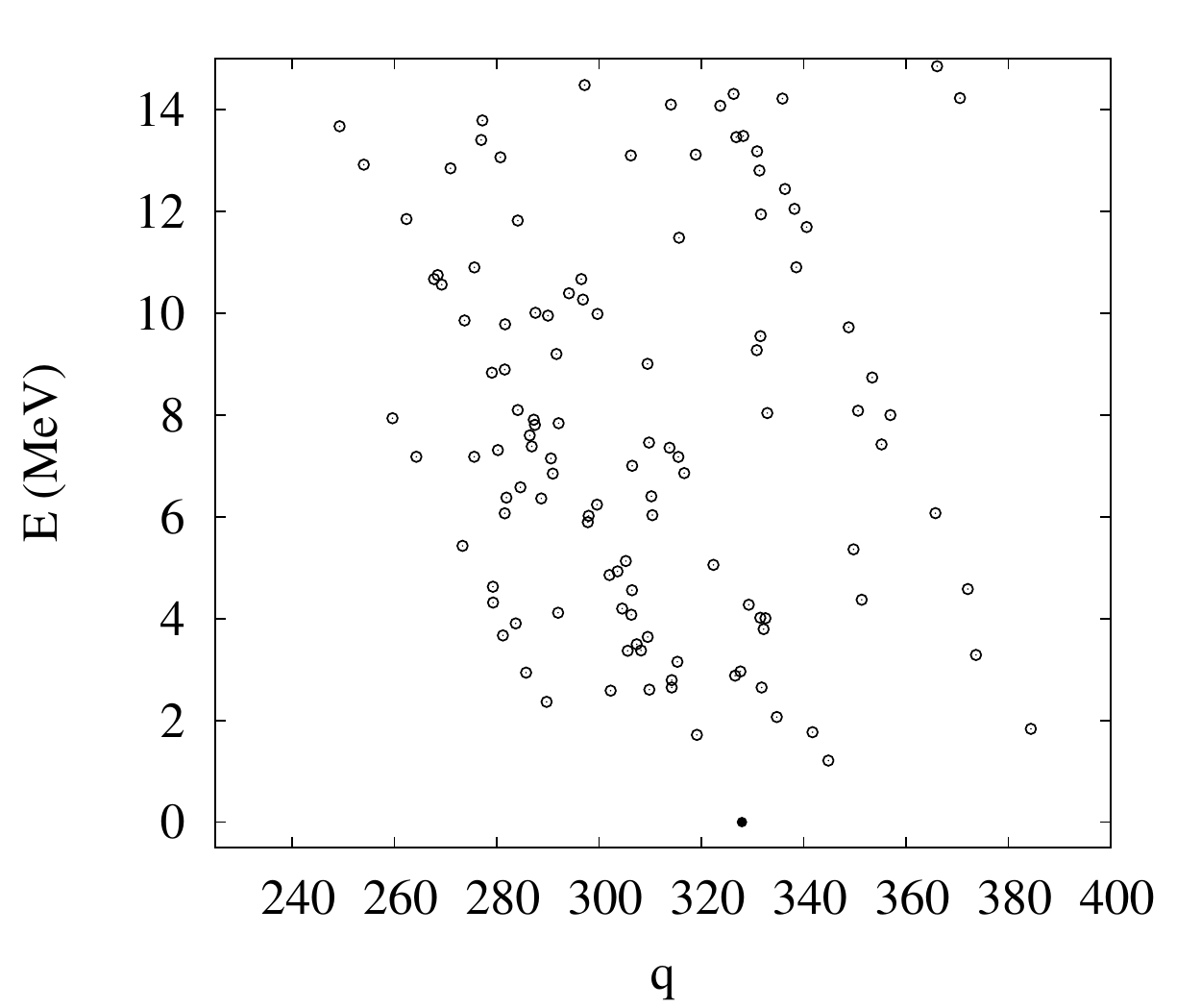}
\caption{\label{dy162}  Ground state configuration of $^{162}$Dy
(filled circle) and excitded state configurations reachable by a single
pair jump (open circles).
}
\end{figure}

The next example exhibits the first steps to building a basis
for treating the dynamics in the configuration-interaction 
framework.  We start with a specific Hamiltonian of the usual
shell-model form Eq. (\ref{H}).
The goal is to examine all possible partitions and determine the
self-consistent minimum of each one.  These states will
span the range of deformations permitted by the shell-model
space.  Additional excited states in the same partitions
may be constructed as particle-hole excitations with the
same mean field as the ground state in that partitions.
We have written a code to carry out the fixed-partition 
ground state minimization, given $H$ of the form Eq. (\ref{H}).
The example we show is the nucleus $^{162}$Dy, see
Fig. \ref{dy162}.
Here we took a Hamiltonian
constructed for use in the Shell Model Monte Carlo 
treatment of level densities~\cite{al08}.  The filled circle shows the
ground state of the axial Hartree-Fock approximation.
This state is moderately deformed, as is expected 
for midrange lanthanide nuclei.  The open circles show the 
lowest excited states that
arise by changing the partition by a single pair jump. One
see a significant gap in the spectrum, and also that the
quadrupole deformations are not very different from that
of the ground state.  The partial localization in deformation
is important to obtain eventually a collective dynamics for
a deformation coordinate.

Now we come to the actinide nuclei and realistic fission paths.  We
start with the two ground states I and II in the PES minima. 
Fig. 8 shows the difference in the partitions of the two states, 
\begin{figure}[tb]
\centering
\includegraphics[width=11cm]{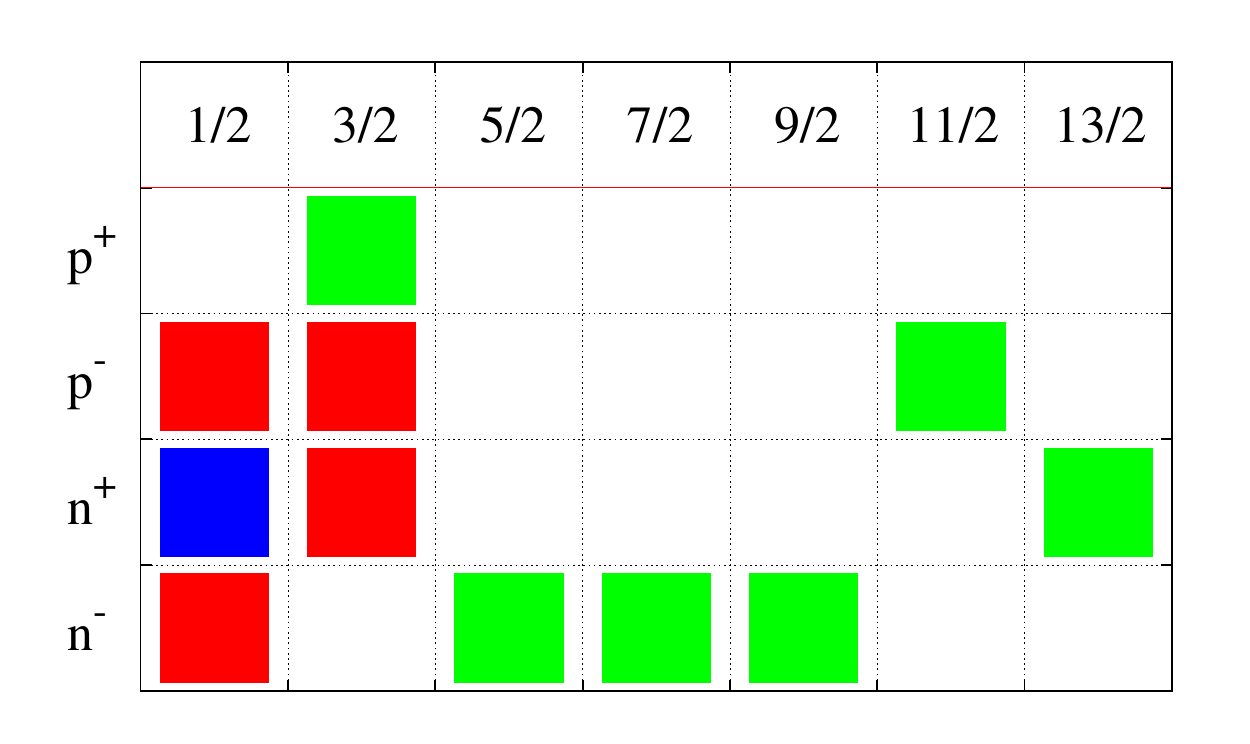}
\caption{\label{fig8}  Difference of partitions in $^{236}$U for the
class I and class II ground states in the FRLD model \cite{mo14}.
}
\end{figure}
as calculated by M\"oller in the FRLD model
\cite{mo95,mo14}.  One sees that particles in higher-$K$ orbitals are
moved to lower $K$ in going from the first minimum to the second.
The number of pairs to be moved totals 6.  To explore the partition
space between the two minima, we should have a self-consistent
mean field code that allows one to specify the partition, with
no other constraints.  Lacking that, we can 
begin to map 
the distance between configurations with existing codes,
using the jump number as the distance measure.  
We did this for $^{236}$U using the HFBaxial
code written by Luis Robledo \cite{robledo_code}.  The configurations
we examined are the I and II ground states, the A and B barriers,
and a number of configurations beyond the second barrier up to
$Q = 140$ bn.  At the largest deformation, the separated fragments have 
the same energy as the elongated fissioning nucleus.  The results
are shown in Fig. \ref{pes-ab}.  First note that the shortest path from
the I ground state to II is not over the saddle. Past the second
saddle, the configurations are labeled by their quadrupole moment.
In this region, the system is asymmetric and parity is not a
good quantum quantum number.   However, for the figure we
have reported jump numbers corresponding to
symmetric fission.  It will be interesting to see the effect
on the jump numbers when the parity constraint is released.
If they remain substantial, it shows that the saddle-to-scission
path requires dynamics beyond that contained in the time-dependent
Hartree-Fock approximation.    
\begin{figure}[tb]
\centering
\includegraphics[width=13cm]{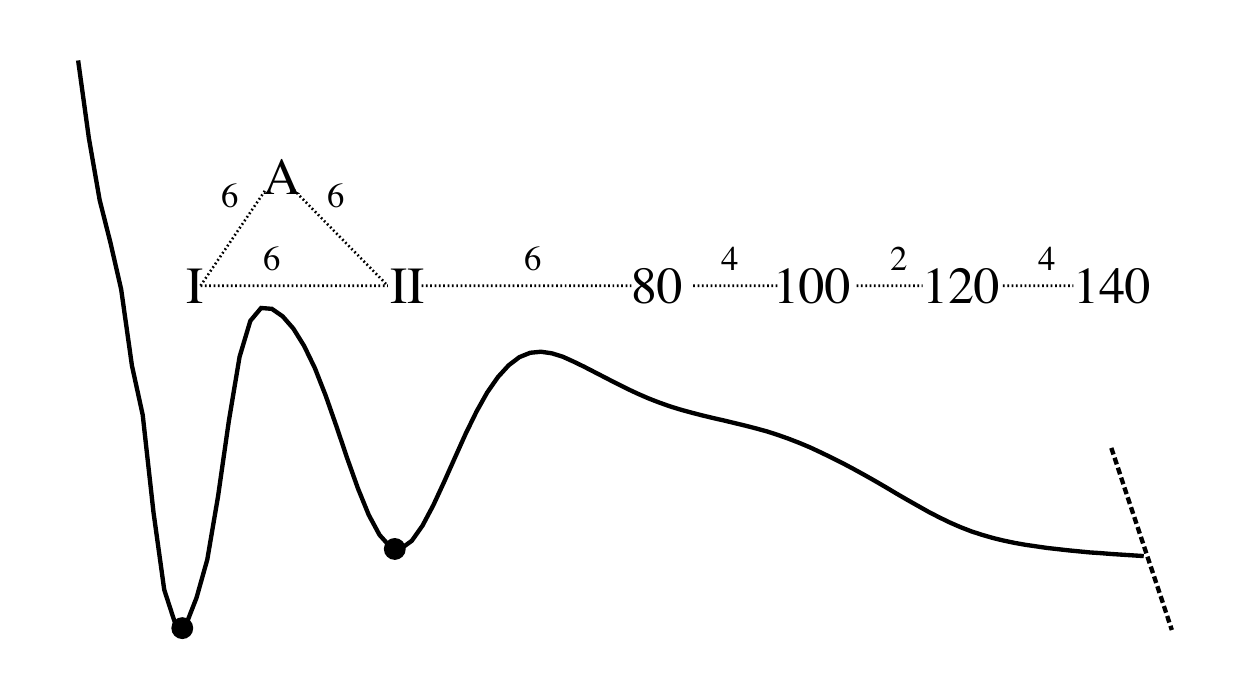}
\caption{\label{pes-ab}
Schematic view of the fission barrier as in Fig. \ref{fig1a}.  The states I,II,A
and B are unconstrained while the states by a quadrupole moment
$Q = 80,100,120,140$ are constrained by that operator.  The distance
between states is shown by the number of pair jumps to get from one to
another.  
}
\end{figure}

\section{The Hamiltonian}
\label{sec-3}   
  While there are many energy density functionals available for
self-consistent mean-field theory, there does not yet exist a
fully self-consistent effective Hamiltonian for nuclear structure.
Still, it should be useful to explore the dynamics that follow
from simplified residual interactions.

\subsection{Interaction between configurations}
\label{sub-1}
The two-particle interaction between configurations can be 
written schematically as 
\be
\label{v-ab}
\langle \alpha | v |\beta\rangle = \sum \langle pp |v | pp \rangle 
\det | \langle \phi^\alpha_i | \phi^\beta_j\rangle|
\ee
where $\langle pp |v | pp \rangle$ is a two-body matrix element and  
$\det | \langle \phi^\alpha_i | \phi^\beta_j\rangle|$
is the overlap of spectator orbitals.  

\begin{figure}[tb]
\centering
\includegraphics[width=11cm]{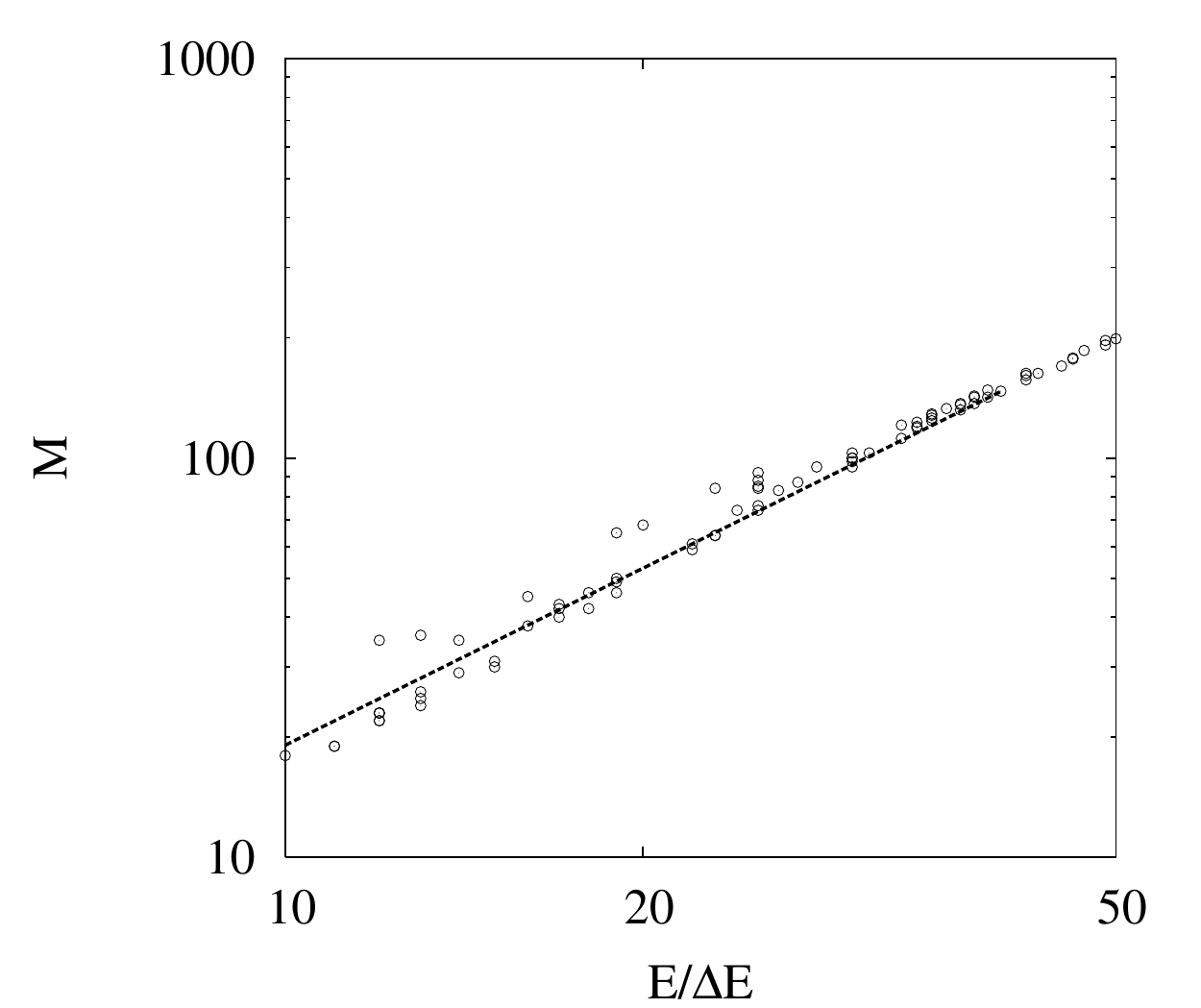}
\caption{\label{fig11}  Relative interaction strengths to mix
configurations at the same energy, as a function of excitation
energy.  See text for explanation.
}
\end{figure}

From the study of the dynamics as seen for example in Fig. \ref{diffusion}, 
it is clear that an important
parameter to approach the diffusion limit
is the ratio of the mean-square average
of the interaction between configurations $\overline{ \langle i| v | j
\rangle^2} $ to the local level density $\rho(q)$.  If this ratio
is large, each state decays exponentially into nearby states and
the dynamics will be diffusive.  It has been
shown in a simple model that this ratio increases with excitation
energy \cite{bu92}, so the diffusion limit will be appropriate at 
high enough excitation energy.  In that model, the average interaction
strength scaled with excitation energy $E$ as
\be
\overline{ \langle \alpha| v | \beta \rangle^2} \sim
E^{3/2}/\rho(E).
\ee
We also have analyzed this with
a somewhat more sophisticated model, with results shown in Fig. \ref{fig11}.
The vertical scale is proportional 
to the product $\overline{ \langle \alpha| v | \beta \rangle^2} 
\rho(E)$.  It is plotted with
respect to excitation energy, measured in units of the single-particle
energy spacing.  The dotted line is a power-law fit to the 
ratios; it comes out very close to the predicted 3/2-power dependence.
Details will be given elsewhere.  

The matrix elements of the two-body interaction are also affected
by the changes in the spectator orbitals due to the different mean
fields of the partitions.  This cuts down the interactions by 
the determinant of spectator orbital overlaps indicated in 
Eq. \ref{v-ab}.
This is a good occasion to mention the early paper by Arima
and Yoshida giving analytic expressions for this determinant in
the harmonic oscillator basis \cite{ar59}.

So where do we hope to go from here?  It seems feasible
in the near term to go beyond simple models
to sample the CI interactions arising from realistic effective
Hamiltonians.  We can then use the derived statistical properties to 
determine where the diffusive limit can be applied.

\section{Acknowledgments}
\label{ack}
We wish to thank A. Bulgac for discussions, L. Robledo for the use of the 
HFBaxial code to calculate partitions of $^{236}$U, and P. M\"oller for 
providing us with the orbital fillings for $^{236}$U in
the FRLD model.  We also acknowledge financial support from
the Institute for Nuclear Theory (GFB) and from the DOE under Grant
FG02-00ER41132 (JMM).


%
%
%

\end{document}